\documentclass{emulateapj}
\usepackage{graphics}

\shorttitle{Swift SNe Ia X-Ray Stacking}
\shortauthors{Russell \& Immler}
\begin{document}

\title{Swift X-Ray Upper Limits on Type Ia Supernova Environments}
\author{B.\ R.\ Russell \altaffilmark{1,2}} \author{S.\ Immler \altaffilmark{2,3,4}}
\altaffiltext{1}{Department of Physics, University of Maryland, College Park, MD 20742. brock@umd.edu}
\altaffiltext{2}{Department of Astronomy, University of Maryland, College Park, MD 20742}
\altaffiltext{3}{Astrophysics Science Division, NASA Goddard Space Flight Center, Greenbelt, MD 20771}
\altaffiltext{4}{Center for Research and Exploration in Space Science and Technology, NASA Goddard Space Flight Center, Greenbelt, MD 20771}

\begin{abstract}
We have considered 53 Type Ia supernovae (SNe Ia) observed by the \textit{Swift} X-Ray Telescope (XRT). None of the SNe Ia are individually detected at any time or in stacked images. Using these data and assuming that the SNe Ia are a homogeneous class of objects, we have calculated upper limits to the X-ray luminosity (0.2 - 10 keV) and mass-loss rate of $L_{\rm 0.2-10} < 1.7 \times 10^{38}$ erg s$^{-1}$ and $\dot{M} < 1.1 \times 10^{-6}$ M$_\odot$ yr$^{-1}\times (v_{\rm w})/(10$ km s$^{-1})$, respectively. The results exclude massive or evolved stars as the companion objects in SNe Ia progenitor systems, but allow the possibility of main sequence or small stars, along with double degenerate systems consisting of two white dwarfs, consistent with results obtained at other wavelengths (e.g., UV, radio) in other studies.
\end{abstract}


\section{Introduction}
SNe Ia are generally considered to be thermonuclear explosions of white dwarfs (WDs) (e.\ g.\ \citealt{Hillebrandt00}). Such explosions can occur if the WD reaches or exceeds the Chandrasekhar limit, although some models suggest that the explosion can occur below the Chandrasekhar limit (\citet{vanKerkwijk}, for example). There are two main classes of models currently considered to be possible progenitor systems for SNe Ia \citep{Hillebrandt00}: 1.) one white dwarf accretes mass from a binary companion until it exceeds the Chandresekhar limit (\citealt{Whelen73,Iben84,Nomoto82}; single-degenerate, SD model) and 2.) two white dwarfs merge (\citealt{Webbink84,Iben84}; double-degenerate, DD model). In the SD model, the companion star emits a stellar wind that populates the circumstellar region with matter, or the star fills the Roche lobe \citep{Branch98}. The SN shock would run into this circumstellar matter and heat it to high enough temperatures ($\sim 10^6 - 10^9$ K) so that it produces thermal X-rays, depending on the fraction accumulated by the white dwarf \citep{Chevalier}. For the DD model both stars in the binary system have long ago ceased producing significant wind, and therefore the circumstellar region will be devoid of matter \citep{Branch98}. Hence, no thermal X-ray emission is expected from shock interaction.

Various observations and studies have been made in an attempt to determine which of these models is correct. In the SD model, there are several methods of mass transfer. One is the so-called ``symbiotic'' binary system, where a fraction of the wind from a binary companion is accreted by the WD until it reaches the Chandrasekhar limit \citep{Panagia06}. \citet{Panagia06} have put constraints on this possibility by analyzing radio observations of a sizable sample of SNe Ia. See below for more discussion of their results. Another possible scenario is that of a massive binary in Roche lobe overflow with material again being accreted onto the WD \citep{Panagia06}. \citet{Panagia06} placed a limit on this, stating that such a method would be required to be 60 -- 70 \% efficient to avoid detectable circumstellar matter (CSM). These limits rely on the conclusions of \citet{Nomoto84} regarding the conditions of accretion to allow a WD to become a SN Ia.

A VLA observation of SN 1986G looking for early radio emission resulting from the interaction of the SN shock with the wind from a red giant companion detected no such emission at 2 cm and 6 cm. The $3\sigma$ upper limits of 0.7 and 1.0 mJy, respectively, are in conflict with those expected from the symbiotic case \citet{Eck95}, although they point out that this was a ``peculiar'' SN Ia. \citet{Panagia06} conducted a radio survey of 27 SNe Ia and detected no radio emission (with a highest radio luminosity upper limit of $4.2 \times 10^{26}$ erg s$^{-1}$), therefore concluding that these systems have low CSM densities. They explicitly argue against the SD model for a massive companion, but state that their results do allow for a relatively low mass companion. They also provide 2$\sigma$ upper limits for the mass loss rate for the system as $3 \times 10^{-8}$ M$_\odot$ yr$^{-1}$, the lowest upper limits published to date.

\citet{Schlegel93} have observed SN 1992A using ROSAT searching for X-rays. They establish an X-ray luminosity upper limit of $3 - 5 \times 10^{38}$ erg s$^{-1}$ and a mass loss rate upper limit of a few $\times 10^{-6}$ M$_\odot$ yr$^{-1}$ \citep{Schlegel93}. \citet{Hughes07} used the Chandra X-ray telescope to observe four SNe Ia. For two of the SNe Ia they observed (SN 2002bo and SN 2005ke), they detected no X-rays and obtained an upper limit of $2 \times 10^{-5}$ M$_\odot$ yr$^{-1}$ for 2002bo. For the other two (SN 2002ic and SN 2005gj), they find upper limits about 4 times lower than would be expected for circumstellar interaction, and propose a mixing scenario to increase X-ray absorption. 

\citet{Immler06} have determined relatively deep upper limits on X-ray emission from SNe Ia using the \textit{Swift} X-ray Telescope (XRT). They examined SN 2005ke, but found inconclusive evidence for X-ray emission. They determine upper limits of luminosity of $(2 \pm 1) \times 10^{38}$ erg s$^{-1}$ and mass loss rate of $3 \times 10^{-6}$ M$_\odot$ yr$^{-1}$ ($v_w/(10$ km s$^{-1})$),  the lowest upper limits published to date from X-ray observations.

\citet{Hancock11} have stacked VLA radio data from 46 observations of nearby SNe Ia to create a deep image, and obtain a radio luminosity upper limit of $1.2 \times 10^{25}$ erg s$^{-1}$ Hz$^{-1}$ at 5 GHz, and a mass-loss rate upper limit for the progenitor system of $1.3 \times 10^{-7}$ M$_\odot$ yr$^{-1}$.

Our goal in this paper is to add to and expand on the previous X-ray studies. Of all past and present X-ray observatories, the \textit{Swift} X-Ray Telescope has observed the largest sample of SNe in X-rays, with unprecedented early observations starting just days after outburst. We use data from XRT to determine an upper limit on X-ray luminosity from more than 50 SNe Ia, adding to the growing consensus regarding the progenitor systems of these events.

\section{Data}
The X-Ray Telescope (XRT; \citealt{XRT}) onboard the \textit{Swift} telescope \citep{Gehrels04} has observed more than 170 SNe to date. Of these, 55 are SNe Ia (at the time of writing). In order to further narrow this sample down to increase the chances of detecting X-rays from the SN, we applied the following conditions for selection: \\ \\
$\bullet$ Young: observations begin within 100 days of SN discovery; \\
$\bullet$ Nearby: distance to SN within 130 Mpc (using previously determined distances from NED and references therein); \\
$\bullet$ Contamination with nearby X-ray source: separation of SN position from nearby X-ray source $>24''$ (corresponding to the XRT 90\% encircled energy width). \\

These conditions lead to the selection of 53 of the SNe Ia that have been observed by the XRT in our calculations. They are listed in Table \ref{LongTable} (which is published in its entirety online), along with their individual luminosity upper limits. Some of these are relatively high due to contamination from the galactic nucleus. All X-ray data used are in the 0.2-10 keV energy band.

\section{Method}
We co-added all individual observations (ObsIDs) using the \textit{ximage} software package (version 4.4.1) to produce a sky image for each supernova. We then used the \textit{ximage} analysis package to determine a $3\sigma$ upper limit on count rate for each supernova image, using a 5 pixel radius corresponding to PSF encircled energy radius of roughly 90\% centered on the optical position of the respective supernova and a background region in a source-free region of the image. These count rates were corrected for 100\% encircled energy radius and were converted to an unabsorbed flux using the online \textit{pimms} tool (version 4.3) using the appropriate column density for the host galaxy \citep{DL} and assuming a 10 keV thermal bremsstrahlung plasma \citep{Fran96}. By making the assumption that all SNe Ia result from the same type of progenitor system, we can follow \citet{Panagia06} to add these 53 upper limits using the equation: $\sigma^2 = \left(\Sigma_i\sigma_i^{-2}\right)^{-1}$, where $\sigma_i$ is the upper limit from the $i$-th supernova and we sum over all of the SNe in our sample.

In order to determine the rate of material lost by Type Ia supernova systems via the winds of a possible companion star, we make several assumptions with respect to model and parameters. We assume spherically symmetric stellar wind, and we calculate the mass loss rate via \citep{ChevFran,Immler06}:
\begin{equation}
L = \frac{1}{4 \pi m^2}\Lambda(T)\left(\frac{\dot{M}}{v_w}\right)^2\left(v_st\right)^{-1}
\end{equation}
where $L_x$ is the X-ray luminosity determined from the data, $m$ is the average particle mass of $1.8 \times 10^{-27}$ kg for a H+He plasma with solar composition, $t$ is the time after outburst determined by weighted average, $v_s$ is the shock speed assumed to be $10\ 000$ km s$^{-1}$, $v_w$ is the wind speed, which for red supergiants ranges from 5 -- 25 km s$^{-1}$, and here we assume it to be $10$ km s$^{-1}$, and $\Lambda$ is the cooling function: $\Lambda = 2.4 \times 10^{-27} g_{ff} T_e^{1/2}$, where $T_e$ is the electron temperature ($T_e = 1.36 \times 10^9 (n-2)^{-2} \left(\frac{v_s}{10^4 km s^{-1}}\right)^2$ K) and $g_{ff}$ is the free-free Gaunt factor. $n$ is the power law index of the SN ejecta: $\rho \propto r^{-n}$ and ranges from 7 - 10. Here we assume it to be 8. The reverse shock is in equipartion unless $T > 5 \times 10 ^8$ K \citep{ChevFran}.

Using the same images, all sources were removed with the detect/remove command in ximage. These `cleaned' images were then analyzed in the same way as above, and results were similar.

\begin{figure}
\plotone{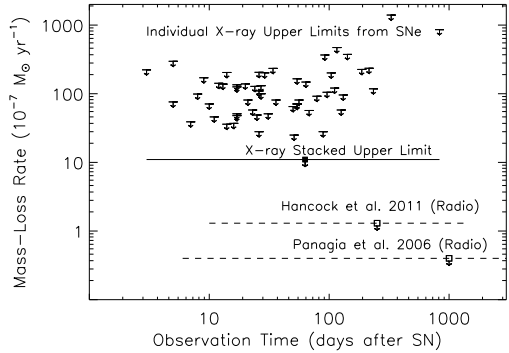}
\caption{
Recent results for mass-loss rate upper limits, including the results from this study and those of \citet{Panagia06} and \citet{Hancock11} (both in radio).
}
\label{fig2}
\end{figure}

\section{Results}
None of the 53 SNe Ia are detected at any time or in stacked images. For the individual SNe, we find the X-ray luminosity $3\sigma$ upper limits range from $2.7 \times 10^{38}$ erg s$^{-1}$ to $2.3 \times 10^{41}$ erg s$^{-1}$ (0.2-10 keV). The mass-loss rate upper limits range from $7.5 \times 10^{-5}$ to $4.0 \times 10^{-6}$ M$_\odot$ yr$^{-1}\times (v_{\rm w})/(10$ km s$^{-1})$. By combining the upper limits for the individual SNe, we obtain a $3 \sigma$ upper limit on the X-ray Luminosity of $1.7 \times 10^{38}$ erg s$^{-1}$ and a $3 \sigma$ upper limit on the mass loss rate of $1.1 \times 10^{-6}$ M$_\odot$ yr$^{-1}\times (v_{\rm w})/(10$ km s$^{-1})$. All luminosities are reported in the 0.2-10 keV energy band.

\section{Discussion}
In this paper we present the deepest flux limits and upper limits on the mass-loss rates of SNe Ia companion objects obtained from X-ray observations to date. These results are combined with those obtained at radio wavelengths of other studies in Figure \ref{fig2}. From these mass-loss upper limits, we can make statements about the progenitor system for a generic, average SN Ia, assuming that SNe Ia form a homogeneous class of objects. 

Radio and X-ray studies of nearby core-collapse SNe have shown that the mass-loss rates of massive progenitors are well above the limit inferred in this study ($10^{-6}$ --$10^{-5}$ M$_\odot$ yr$^{-1}$). Wind accretion is generally about 10\% efficient \citep{Y}. If 10\% of the wind is being accreted onto the white dwarf, that would increase the mass loss from the companion by 10\% over the upper limits presented here. The result is still below the mass-loss expected from a massive star, and therefore we can rule out wind from a massive star accreting onto a white dwarf as the progenitor system for SNe Ia. We cannot rule out a main sequence star as a companion with mass-loss rates $<10^{-7}$ M$_\odot$ yr$^{-1}$ in a wind-accretion scenario. A double-degenerate white dwarf system is also not excluded due to the lack of CSM interaction and hence the lack of emitted X-rays. The possibility of a massive companion in Roche Lobe overflow is also still permitted by these limits. Our results qualitatively support results from other wavelength regimes, such as in the radio \citep{Panagia06,Hancock11} and in the UV \citep{B}.

Our results improve on the previous lowest published upper limit derived from X-ray observations, $3 \times 10^{-6}$ M$_\odot$ yr$^{-1}$ $v_w$/(10 km s$^{-1}$) from \citet{Immler06}. This new limit is about 1/3 that of theirs.

We also compare our results to those of \citet{Sternberg}, who had detected blueshifted sodium spectra from SNe Ia. Our list overlaps with their blueshifted results in only two supernovae: 2006X and 2009ig. We do not detect X-rays from these SNe, and the mass-loss rate upper limit we determine for these two are $3.2 \times 10^{-6}$ and $4.2 \times 10^{-6}$ M$_\odot$ yr$^{-1}$ respectively.

Since the X-ray emission from Type Ia Supernovae results from a different process than radio emission and from different regions in the shocked circumstellar medium, these results provide a completely independent check on the results of others.

The mass-loss rate is sensitive to both time after outburst of observation and distance of the supernova. Therefore, it is possible for individual SNe to give similar or even lower upper limits than the stacked data; the stacked data, however, give results for the entire sample as a whole. These results give upper limits for a homogeneous class of objects, but it is not certain whether all SNe Ia have the same progenitor systems.

Several of the observations included in this study have been of SNe at very early times (for example, SN 2005bc was observed at 2.7 days after discovery of the SN), when the X-ray emission would be highest due to the higher CSM density found at smaller radii. For these early observations, the upper limit we derive for the mass-loss rate is on the order of $10^{-5}$ M$_\odot$ yr$^{-1}$ $v_w$/10 km s$^{-1}$. The X-ray luminosity upper limits are on the order of $10^{41}$ erg s$^{-1}$.

Further early and sensitive observations are still needed to address questions about the nature of SNe Ia progenitor systems. Prompt X-ray observations during or shortly after the explosion will enable us to probe the circumstellar environment down to the surface of the white dwarf, allowing us to further constrain the properties of the companion star.

\section{Summary}
Using the largest sample of SNe Ia available for stacking analysis, we have generated a deep \textit{Swift} XRT X-ray image of 3.05 Ms exposure time. We determine that there is no X-ray source at the position of the stacked SNe. We calculate a $3\sigma$ upper limit for the X-ray luminosity of $1.7 \times 10^{38}$ erg s$^{-1}$, and an upper limit on the mass-loss rate of $1.1 \times 10^{-6}$ M$_\odot$ yr$^{-1}\times (v_{\rm w})/(10$ km s$^{-1})$ from the SN progenitors. The low upper limits provide further evidence that the companion stars in SNe Ia progenitor systems are not a massive star (e.g. red supergiant or post main-sequence) in a wind-accretion scenario. Lower-mass main sequence stars with small amounts of mass lost in stellar winds, a white dwarf companion in double-degenerate SNe Ia systems, or a massive star in Roche lobe overflow are not excluded. It is also possible that SNe Ia do not form a homogeneous class of objects, in which case there may be individual detections of X-rays from some SNe Ia.

\vfill

\clearpage 

\begin{deluxetable}{lcccccc}
\tablecaption{Type Ia Supernova Properties \label{IaTable}}
\tablewidth{0pt}
\tablehead{
\colhead{} & \colhead{Weighted} & \colhead{} & \colhead{} & \colhead{} & \colhead{Luminosity} & \colhead{Mass Loss} \\
\colhead{} & \colhead{Average} & \colhead{} & \colhead{} & \colhead{Exposure} & \colhead{(0.2-10 keV)} & \colhead{Rate} \\
\colhead{Supernova} & \colhead{Date} & 
\colhead{Date Range} & \colhead{Distance} & \colhead{Time} & 
\colhead{$3\sigma$ upper limit} & \colhead{$3\sigma$ upper limit} \\
\colhead{} & \multicolumn{2}{c}{[days after discovery]} & \colhead{[Mpc]} & \colhead{[ks]} &
 \colhead{[$10^{38}$ erg s$^{-1}$]} & \colhead{[$\frac{10^{-6} \mathrm{M_\odot yr}^{-1}}{10 \mathrm{km s}^{-1}}$]}}
\startdata 
2005am	&	55.9	&	11.0	--	84	&	30	&	71.7	&	20.2	&	9.0	\\
2005bc	&	2.7	&	2.7	--	2.7	&	52	&	0.7	&	2184.2	&	20.7	\\
2005cf	&	53.7	&	7.0	--	462.4	&	29	&	66.4	&	14.5	&	7.5	\\
2005df	&	233.5	&	6.1	--	860.5	&	16	&	50.7	&	4.4	&	8.5	\\
2005gj	&	831.7	&	59.1	--	1257.1	&	50	&	14.7	&	94.7	&	75.0	\\
2005hk	&	187.0	&	4.8	--	935.1	&	56	&	57.7	&	39.3	&	22.9	\\
2005ke	&	50.8	&	1.7	--	1228.8	&	19	&	288.3	&	2.7	&	3.2	\\
2006dd	&	131.0	&	1.8	--	120.2	&	23	&	100.8	&	199.9	&	43.2	\\
2006dm	&	142.2	&	2.6	--	728.6	&	87	&	80	&	145.7	&	38.4	\\
2006E	&	126.3	&	1.8	--	351.0	&	11	&	31.8	&	4.4	&	6.3	\\
2006ej	&	91.6	&	1.8	--	490.7	&	78	&	48.3	&	147.5	&	31.0	\\
2006mr	&	54.2	&	2.6	--	172.0	&	23	&	56.6	&	238.3	&	30.3	\\
2006X	&	26.1	&	4.3	--	668.6	&	17	&	58.9	&	5.4	&	3.2	\\
2007af	&	68.1	&	1.8	--	288.0	&	25	&	125.6	&	10.3	&	7.1	\\
2007ax	&	96.3	&	8.8	--	576.0	&	31	&	65.1	&	67.2	&	21.5	\\
2007bm	&	214.9	&	6.7	--	578.0	&	26.3	&	28.4	&	38.5	&	24.3	\\
2007co	&	115.9	&	5.8	--	386.3	&	99	&	54.8	&	276.7	&	47.8	\\
2007cq	&	104.1	&	4.7	--	390.4	&	50	&	53.3	&	53.3	&	19.9	\\
2007cv	&	111.4	&	1.8	--	404.0	&	32.1	&	62.9	&	28.3	&	15.0	\\
2007gi	&	6.7	&	3.6	--	9.2	&	20.4	&	11.1	&	32.9	&	4.0	\\
2007on	&	49.7	&	1.7	--	416.6	&	28	&	108.1	&	36.1	&	13.2	\\
2007S	&	64.4	&	7.6	--	312.7	&	60	&	83.7	&	83.2	&	19.6	\\
2007sr	&	30.6	&	2.7	--	192.1	&	28	&	92.4	&	12.7	&	5.3	\\
2008A	&	53.5	&	5.6	--	318.4	&	70	&	51	&	89.5	&	18.5	\\
2008ae	&	9.26	&	2.7	--	873.5	&	127	&	18.8	&	434	&	16.9	\\
2008dx	&	5.4	&	3.9	--	7.1	&	97	&	5.4	&	2328.2	&	29.8	\\
2008ge	&	23.2	&	14.7	--	28.8	&	16	&	8.3	&	19.3	&	5.6	\\
2008ha	&	36.4	&	4.3	--	88.0	&	22.5	&	11	&	26.8	&	8.3	\\
2008hs	&	25.9	&	2.0	--	58.6	&	74.2	&	68	&	123.2	&	15.1	\\
2008hv	&	89.4	&	2.7	--	60.3	&	48.8	&	63.5	&	28.2	&	13.4	\\
2008Q	&	78.9	&	5.0	--	494.1	&	32	&	64.2	&	16	&	9.5	\\
2009an	&	10.9	&	4.3	--	25.8	&	39.9	&	60.7	&	27.6	&	4.6	\\
2009cz	&	24.3	&	8.7	--	433.8	&	90	&	72.9	&	93.7	&	12.7	\\
2009dc	&	29.5	&	16.5	--	54.2	&	89.3	&	28.8	&	198.9	&	20.5	\\
2009gf	&	26.9	&	1.7	--	68.0	&	75.8	&	70	&	46.4	&	9.4	\\
2009ig	&	16.3	&	1.8	--	53.4	&	36.7	&	132	&	14.2	&	4.1	\\
2009iz	&	5.2	&	3.7	--	8.0	&	55	&	11.5	&	160.4	&	7.7	\\
2009jr	&	12.4	&	7.7	--	17.4	&	64	&	18.8	&	227.3	&	14.2	\\
2009Y	&	16.5	&	4.8	--	55.4	&	123.7	&	136.9	&	735.1	&	29.4	\\
2010ae	&	17.4	&	3.6	--	33.2	&	17	&	11.2	&	22.5	&	5.3	\\
2010el	&	9.5	&	4.1	--	15.8	&	30.19	&	6.5	&	585.4	&	19.9	\\
2010ev	&	13.6	&	3.9	--	32.2	&	32.25	&	75.7	&	38.2	&	6.1	\\
2010fy	&	7.7	&	3.7	--	11.5	&	50	&	10.7	&	180.7	&	10.0	\\
2010gp	&	33.6	&	10.6	--	59.0	&	100	&	49.8	&	260.7	&	25.0	\\
2010hh	&	14.5	&	2.1	--	12.2	&	80	&	8.8	&	408.9	&	20.6	\\
2010Y	&	16.8	&	1.9	--	36.1	&	46	&	8.1	&	398.1	&	21.9	\\
2010ih	&	16.7	&	6.6	--	27.8	&	17.05	&	17.1	&	19.1	&	4.8	\\
2010kg	&	19.9	&	4.0	--	36.6	&	19.24	&	43.4	&	136.8	&	13.9	\\
2010ko	&	20.6	&	2.0	--	33.3	&	10.78	&	59.8	&	44.9	&	8.1	\\
2011B	&	24.7	&	4.7	--	79.0	&	25.5	&	74.7	&	13.1	&	4.8	\\
2011M	&	13.2	&	6.6	--	20.0	&	13.38	&	35.3	&	349.6	&	18.1	\\
2011aa	&	26.7	&	5.6	--	59.6	&	25.35	&	56.1	&	52.4	&	10.0	\\
2011ao  &       16.8    &                               &       36.8    &       93.0    &       17.1    &       4.5
\enddata	
\end{deluxetable}

\end{document}